\documentclass[pra,twocolumn,byrevtex,showpacs]{revtex4}

\usepackage{graphicx}
\usepackage{dcolumn}
\usepackage{amsmath}
\usepackage{epsfig}
\usepackage{bm}
\usepackage{amssymb}
\usepackage{sidecap}

\usepackage[dvips]{hyperref}
\hypersetup{colorlinks=true,linkcolor=blue,citecolor=blue,urlcolor=black}

\begin{document}

\title{Measurement strategy for spatially encoded photonic qubits}
\date{\today}
\author{M. A. Sol\'is-Prosser}
\author{L. Neves}
\email{leonardo.neves@cefop.udec.cl} 

\affiliation{Center for Optics and Photonics, Universidad de Concepci\'on, Casilla 4016, Concepci\'on, Chile}
\affiliation{Departamento de F\'isica, Universidad de Concepci\'on, Casilla 160-C, Concepci\'on, Chile}



\begin{abstract}  
We propose a measurement strategy which can, probabilistically, reproduce the statistics of \emph{any} observable for spatially encoded photonic qubits. It comprises the implementation of a two-outcome positive operator-valued measure followed by a detection in a \emph{fixed} transverse position, making the displacement of the detection system unnecessary, unlike previous methods. This strategy generalizes a scheme recently demonstrated by one of us and co-workers restricted to measurement of observables with equatorial eigenvectors only. The method presented here can be implemented with the current technology of programmable multipixel liquid-crystal displays. In addition, it can be straightforwardly extended to high-dimensional qudits and may  be a valuable tool in optical implementations of quantum information protocols with spatial qubits and qudits.
\end{abstract}

\maketitle

\section{Introduction}
Optical implementations of quantum information technologies based on the spatial degree of freedom of single photons have at their disposal several approaches to embody a qubit or qudit structure into this continuous variable. For instance, one can encode the qubit in the spatial parity of the photon's one-dimensional transverse modes \cite{Abouraddy07} or one can truncate (via filtering) the infinite-dimensional discrete Hilbert space of Laguerre-Gaussian \cite{Mair01}  or Hermite-Gaussian \cite{Walborn05} modes in which the photon's two-dimensional transverse distribution may be expanded. In simpler approaches, however, the qubits and qudits can be encoded either in the discretized transverse position--momentum of the photons \cite{Neves04,Neves05,Sullivan-Hale05} or their angular-position--orbital-angular-momentum variables \cite{Jha10}. Here, we consider this former type of spatial encoding where the discretization of the one-dimensional transverse modes is achieved by making the photons pass through an aperture with $D$ slits [e.g., see Fig.~\ref{fig:qudit_transform}(a) for $D=2$] \cite{Neves04,Neves05}. Hereafter we will refer to them as spatial qubits and qudits.

Recently, considerable effort has been directed toward finding methods to measure spatial qubits and qudits. This is a fundamental issue for any possible application since the extraction of information from a quantum system requires the performance of a  measurement. In this respect, Neves \emph{et al.} \cite{Neves07} have shown that the spatial filtering in transverse positions at the near and far field of a double slit corresponds to projections in the poles and equator of the Bloch sphere [Fig.~\ref{fig:qudit_transform}(b)], respectively. This enabled characterization of the entanglement \cite{Neves07,Peeters09} and quantum tomography \cite{Lima08,Taguchi08} of two-photon spatial qubits. Later,  Taguchi \emph{et al.} \cite{Taguchi08} went a step further and showed that any vector over the Bloch sphere can be measured by moving  a ``pointlike'' detector in a transverse position between the near field and the far field of the double slit, both achieved with the help of a lens system [Fig.~\ref{fig:qudit_transform}(a)]. Despite successfully accessing the whole Bloch sphere, this method is difficult and time consuming in practice since it requires moving the detection system for each measurement. Also, the detection efficiency decreases when the detector moves away from the diffraction envelope peak(s), which demands artificial compensation in the count rates when a pair of basis vectors is measured in asymmetric positions around the optical axis.

A different approach for measuring only the \emph{equatorial} states of spatial qubits and qudits has been recently developed by one of us and co-workers \cite{Lima10a}. There, phase shifts were imprinted in each slit of the array (by a spatial light modulator) and after this the photon was detected in a fixed transverse position in the focal plane of a lens, corresponding to an equally weighted superposition of the ``which-slit'' states without any relative phase. This enabled the postselection of noncanonical mutually unbiased basis (MUB) and the realization of the so-called MUB tomography \cite{Lima10a,Lima10b}. 

In this work, we generalize on this latter scheme and propose a measurement strategy which can, probabilistically, reproduce the statistics of \emph{any} observables, not only those with equatorial eigenvectors. The strategy is as follows: First, we implement a two-outcome positive operator-valued measure (POVM), which, with a given success probability, transforms the spatial qubit state into a convenient form. Then, the successfully transformed qubit is detected in a fixed transverse position. By recording and normalizing the count rates for two such measurements, corresponding to two orthogonal projectors of \cite{Taguchi08}, we can reproduce the statistics of the observable. Thus, the method presented here requires neither moving the detection system nor compensation in the count rates and can be implemented with the current technology of multipixel liquid-crystal displays. In addition, it can be straightforwardly extended to higher dimensions and may  be a valuable tool in optical implementations of quantum information protocols with spatial qubits and qudits.

\section{Spatial postselection}  \label{sec_japa}
\begin{figure}[t]
\centerline{\includegraphics[width=0.48\textwidth]{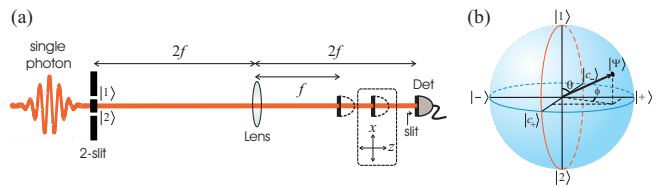}}
\caption{\label{fig:qudit_transform}  (Color online) (a) Schematic representation of the method to measure spatial qubits reported in \cite{Taguchi08}.  (b) Bloch sphere: The great circle (in red) connecting $|1\rangle,|2\rangle$, and $|c_\pm\rangle$ represents the only measurements which do not require compensation (see text for details).}
\end{figure}

Assume that a source produces a monochromatic single-photon field with an arbitrary spatial shape. If the photon is made to pass through a double slit [as shown in Fig.~\ref{fig:qudit_transform}(a)], its quantum state right after it will be given by a $2\times 2$ density matrix, $\hat{\rho}$, which can be pure or mixed depending on the coherence properties of the source and the relevant dimensions of the double slit \cite{Neves04,Neves05,Neves07,Peeters09,Lima08,Taguchi08}.  Typically,  $\hat{\rho}$ is written in the basis $\{|1\rangle,|2\rangle\}$, which represents the slit where the photon was transmitted, so that
\begin{equation}  \label{rho_spa_qb}
\hat{\rho} = \sum_{i,j=1}^2\rho_{ij}|i\rangle\langle j|,
\end{equation}
where $\mathrm{Tr}(\hat{\rho})=1$, $\hat{\rho}=\hat{\rho}^\dagger$, and $\langle \psi|\hat{\rho}|\psi\rangle\geq 0, \forall\; |\psi\rangle$.

Taguchi \emph{et al.} \cite{Taguchi08} showed that when the photon propagates through a lens system and is detected by varying the transverse position ($x$) of a pointlike detector between its focal plane (far field) and image plane (near field), one can access the whole Bloch sphere surface of the spatial qubit [Fig.~\ref{fig:qudit_transform}(b)]. Figure~\ref{fig:qudit_transform}(a) shows a schematic representation of this method. A lens of focal length $f$ is placed at a distance $2f$ from the double slit. The pointlike detector can be displaced in the transverse and longitudinal directions, $x$ and $z$, respectively, where $x\in(-\infty,+\infty)$, in principle, and $z\in[f,2f]$. The effect of a measurement over the state $\hat{\rho}$ in Eq.~(\ref{rho_spa_qb}) for a given position $(x,z)$ is to \emph{postselect} it in a non-normalized state given by 
\begin{equation}  \label{spa_projec}
|\varphi(x,z)\rangle = \sum_{j=1}^2\varphi_j(x,z)|j\rangle ,
\end{equation}
which, after normalization, corresponds to a point on the Bloch sphere surface characterized by spherical polar coordinates $\theta$ and $\phi$, as shown in Fig.~\ref{fig:qudit_transform}(b). The wave function associated with each slit ($j=1,2$) is given by
\begin{equation}  \label{amp_projec}
\varphi_j(x,z) = \sqrt{\frac{\kappa}{\pi}}e^{i\kappa xd(j-1)/a}\mathrm{sinc}\left\{\kappa\left[ x+d(j-\Delta)\eta\right]\right\},
\end{equation}
where $\Delta=(D+1)/2$, $d$ is the center-to-center separation between the slits, $a$ is the slit half-width and $\mathrm{sinc}(\epsilon)\equiv(\sin\epsilon)/\epsilon$. Further parameters are $\kappa=2\pi a/\lambda\eta Z$, $\eta=(z-f)/f$, and $Z=(2f-z)/\eta$ \cite{Taguchi08}. 

By introducing the operator $\hat{S}(x,z)=|\varphi(x,z)\rangle\langle\varphi(x,z)|$, the detection probability \emph{density} for a photon in the state (\ref{rho_spa_qb}) at the position ($x,z$) becomes (under the assumption that the detector has unit quantum efficiency) $\mathcal{P}_\mathrm{spa}(x,z) =  \mathrm{Tr}[\hat{\rho}\hat{S}(x,z)]$. This quantity does not give, directly, information about observable quantities for the spatial qubits. In order to measure observables here, one must determine the longitudinal plane $z$ and two transverse positions $x_\ell$ ($\ell=1,2$) such that the vectors in Eq.~(\ref{spa_projec}) are orthogonal, and are thereby the eigenstates of the observable. Then, the normalized count rates for these two positions give the probabilities for the two possible outcomes labeled by $\ell=1,2$ \cite{Neves07,Peeters09,Lima08,Taguchi08}. Mathematically, this probability is written as 
\begin{equation}
P_\mathrm{spa}^{(\ell)} = \sum_{i,j=1}^2\rho_{ij}\varphi_{i\ell}^*\varphi_{j\ell},
\label{prob_spa}
\end{equation}
where $\varphi_{j\ell}=\varphi_j(x_\ell,z)/\sqrt{\sum_{i=1}^2|\varphi_i(x_\ell,z)|^2}$ and $P_\mathrm{spa}^{(1)}+P_\mathrm{spa}^{(2)}=1$. As an example, at the image plane ($z=2f$), the Pauli operator $\hat{\sigma}_z$ is measured by detecting in the positions $x_\ell=(-1)^{\ell}d/2$, corresponding to $|\ell\rangle\langle\ell|$. 

As mentioned before, this method can be difficult and time consuming since it requires the displacement and accurate positioning of the detection system for each measurement. In addition, when a pair of basis vectors is measured in asymmetric positions around the optical axis, the detection efficiency between them is different due to the diffraction envelope. One must then compensate for this by using the theoretically expected ratios between the corresponding values of the diffraction envelope at those positions. The measurements of observables whose eigenvectors lie in the great circle connecting $|1\rangle,|2\rangle$ and $|c_\pm\rangle$ [red (vertical) circle in Fig.~\ref{fig:qudit_transform}(b)] are the \emph{only} ones which do not require compensation.

\section{Alternative measurement strategy}  \label{sec_POVM}
Here we present an alternative strategy for measuring observables for spatial qubits, which is divided into two steps.

\subsection{The POVM and its physical implementation}
The first step of our strategy is the implementation of a two-outcome POVM whose goal is to transform, probabilistically, the initial state of the spatial qubit into a convenient form which will be discussed in the second step. The physical implementation of a POVM requires the extension of the Hilbert space of the system to be measured. This can be provided by an ancillary quantum system, or \emph{ancilla} \cite{BarnettBook}. Here, the ancilla for the spatial qubit will be the photon polarization.  Therefore, the Hilbert space of this larger system is $\mathcal{H}_{\mathrm{spa}}\otimes\mathcal{H}_{\mathrm{pol}}$, where $\mathcal{H}_{\mathrm{spa}}$($\mathcal{H}_{\mathrm{pol}}$) is the Hilbert space of the spatial (polarization) qubit. Assuming that the systems are initially independent, we start by preparing the ancilla in a known state, say with horizontal polarization $|H\rangle$. Then, we let them evolve subjected to a controlled interaction given by a unitary operator $\hat{U}$ acting on $\mathcal{H}_{\mathrm{spa}}\otimes\mathcal{H}_{\mathrm{pol}}$. This interaction entangles the spatial and polarization degrees of freedom and, after it, we measure the ancilla in the basis $\{|H\rangle,|V\rangle\}$ (where $|V\rangle$ denotes vertical polarization). The POVM will then emerge as the residual effect on the spatial qubit due to the entanglement created by $\hat{U}$. Finally, we evaluate the postmeasurement state of the spatial qubit and the probability for each of the two possible outcomes.

Let $\hat{\Pi}_p$ be the elements of the POVM just described, where the subscript $p=H,V$ labels the two possible outcomes. Mathematically, they can be written as
\begin{eqnarray}   \label{POVM_el}
\hat{\Pi}_p & = & \hat{A}_p^\dagger\hat{A}_p \nonumber\\
&=& \langle H|\hat{U}^\dagger|p\rangle\langle p|\hat{U}|H\rangle .
\end{eqnarray}
One can easily check that all properties of a POVM are satisfied: hermiticity, $\hat{\Pi}^{\dagger}_p=\hat{\Pi}_p$; positivity, $\langle\psi|\hat{\Pi}_p|\psi\rangle\geq 0$, $\forall\;|\psi\rangle$; and completeness, $\sum_p\hat{\Pi}_p=\hat{I}_{\mathrm{spa}}$, where $\hat{I}_{\mathrm{spa}}$ represents the identity on $\mathcal{H}_{\mathrm{spa}}$. For our purposes the unitary operator will be defined as
\begin{equation}  \label{unitary}
\hat{U}=\sum_{j=1}^{2}e^{i\phi_j}|j\rangle\langle j|\otimes\hat{R}(\theta_j),
\end{equation}
where
\begin{equation}  \label{rotation}
\hat{R}(\theta_j)= 
\left(\begin{array}{cc}
\cos(2\theta_j) & -\sin(2\theta_j) \\[2mm]
\sin(2\theta_j) & \cos(2\theta_j)
\end{array}\right).
\end{equation}
The operator $\hat{R}(\theta_j)$, acting on $\mathcal{H}_{\mathrm{pol}}$, is written in the $\{|H\rangle,|V\rangle\}$ basis. It is easy to see that $\hat{U}\hat{U}^\dagger=\hat{U}^\dagger\hat{U}=\hat{I}$, where $\hat{I}$ is the identity on $\mathcal{H}_{\mathrm{spa}}\otimes\mathcal{H}_{\mathrm{pol}}$. This unitary operation rotates the photon polarization by $2\theta_j$ and adds a phase shift $\phi_j$, conditional to the passage of the photon through the slit $j$. Therefore, after $\hat{U}$, spatial and polarization degrees of freedom become entangled and the projective measurement $|p\rangle\langle p|$ on the ancilla polarization accomplishes the POVM on the spatial qubit.

If the state of the spatial qubit prior to the measurement is $\hat{\rho}$ and the outcome of the measurement is $p$, then the postmeasurement state of the system is $\hat{\rho}_p = \hat{A}_p^\dagger\hat{\rho}\hat{A}_p/\mathrm{Tr}(\hat{\Pi}_p\hat{\rho}).$
The denominator gives the probability ($P_p$) for the outcome $p$, or equivalently, the probability that the state $\hat{\rho}_p$ has been prepared. Explicitly, if $p=H$, the postmeasurement state using Eqs.~(\ref{rho_spa_qb}) and (\ref{POVM_el})--(\ref{rotation}), will be given by (omitting the polarization)
 \begin{equation}  \label{rho_H}
\hat{\rho}_H = \frac{1}{P_H}\sum_{i,j=1}^{2}\rho_{ij}e^{i(\phi_i-\phi_j)}\cos(2\theta_i)  \cos(2\theta_j) |i\rangle\langle j|,
\end{equation}
and it is prepared with probability  
\begin{equation}
P_H = \mathrm{Tr}(\hat{\Pi}_H\hat{\rho}) .
\label{P_H}
\end{equation}

\begin{figure}[t]
\centerline{\includegraphics[width=0.48\textwidth]{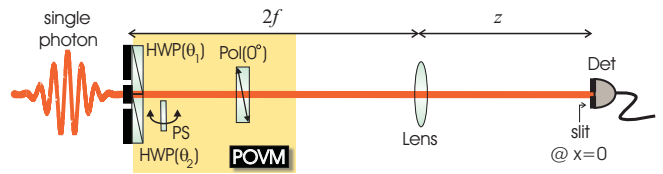}}
\caption{\label{fig:POVM}  (Color online) Schematic representation of the method to measure spatial qubits proposed in this work. HWP: half-wave plate; PS: phase shifter; Pol: polarizer that transmits $H$-polarized light.}
\end{figure}

An experimentally feasible way of realizing this POVM is sketched in the box shown in Fig.~\ref{fig:POVM}. First, the photon incident on the double slit is prepared with polarization $|H\rangle$, so that after transmission its quantum state will be $\hat{\rho}\otimes|H\rangle\langle H|$ with $\hat{\rho}$ given by Eq.~(\ref{rho_spa_qb}). Second, a rotating half-wave plate (HWP) behind each slit of the double slit, followed by a phase shifter (PS) in one of the slits, implements the conditional unitary operation given by Eqs.~(\ref{unitary}) and (\ref{rotation}), which couples spatial and polarization degrees of freedom. Finally, the polarization is measured in the basis $\{|H\rangle,|V\rangle\}$ with a polarizer \cite{Comment2}. Clearly, to build such a  device in a homemade fashion is difficult in practice given the very small dimensions ($a$ and $d$) of the apertures (usually, tens of micrometers). However it can be easily implemented with the help of programmable liquid-crystal displays (LCDs). LCDs are multipixel optical devices which can, in a controlled way, rotate the light polarization according to the pixel's variable phase retardance. Recently, they have been used to manipulate the amplitude \cite{Lima09} and phase \cite{Lima10a,Lima10b} of spatial qudits with high accuracy, which ensures the feasibility of the POVM described here, even for high-dimensional qudits as we discuss later. 

\subsection{Photon detection in a fixed transverse position}
The second step of our strategy comprises the photon detection in the position $(0,z)$  \cite{Comment} \emph{if} the outcome of the previous measurement was $p=H$ (see Fig.~\ref{fig:POVM}). Therefore, it will be realized with probability $P_H$ given by Eq.~(\ref{P_H}), which makes the whole process probabilistic. In this case, the spatial qubit state, which has been transformed into $\hat{\rho}_H$ [Eq.~(\ref{rho_H})], is postselected in the non-normalized state $|1\rangle+|2\rangle$, according to Eqs.~(\ref{spa_projec}) and (\ref{amp_projec}). The \emph{total} probability density for detecting the photon will be 
\begin{eqnarray}
\mathcal{P}_{\mathrm{tot}}(0,z) &=& P_{H}\mathcal{P}_{\mathrm{spa}}(0,z) \nonumber\\
&=& |\varphi_1(0,z)|^2\sum_{i,j=1}^{2}\rho_{ij}e^{i\phi_i}\cos(2\theta_i)
 \nonumber \\
& & \text{} \times e^{-i\phi_j}\cos(2\theta_j)  ,
\label{density_p}
\end{eqnarray}
where $\varphi_1(0,z)=\varphi_2(0,z)$ and is given by Eq.~(\ref{amp_projec}).

By looking at Eqs.~(\ref{prob_spa})  and (\ref{density_p}) it is easy to check that if we set the HWP angles and the phase shift obeying the conditions
$$\theta_{1}\equiv\theta_\ell=\frac{1}{2}\cos^{-1}(|\varphi_{1\ell}|), \;\;\;\;\;\;\;
\theta_{2} = \theta_{\ell}-\frac{\pi}{4}, $$
\begin{equation}
\phi_{2}-\phi_{1}\equiv\phi_\ell=\arg\left(\frac{\varphi_{1\ell}}{\varphi_{2\ell}}\right), \label{angles_cond}
\end{equation}
the \emph{sum} in those equations will be identical. Here $\ell$ labels a possible set of angles $(\theta_\ell,\phi_\ell)$. Hence, by choosing two sets ($\ell=1,2$) corresponding to two orthogonal spatial projectors, we will be able to reproduce the statistics of the observable  associated with these basis vectors. Denoting the probability for each outcome as $P^{(\ell)}_{\mathrm{tot}}$, we get
\begin{equation}  \label{prob_equal}
P^{(\ell)}_{\mathrm{tot}}=P^{(\ell)}_{\mathrm{spa}},
\end{equation}
such that $P^{(1)}_{\mathrm{tot}}+P^{(2)}_{\mathrm{tot}}=1$. Experimentally, this will be achieved by recording the count rate for each set $(\theta_\ell,\phi_\ell)$ and then normalizing the obtained data. 

The role of the POVM $\{\hat{\Pi}_p\}$ (\ref{POVM_el}) in the context of measurement of observables is now clear. It transforms the initial state (\ref{rho_spa_qb}) into state (\ref{rho_H}) with the angles given by Eq. (\ref{angles_cond}), such that its new matrix elements are (up to normalization $P_H$) identical to those elements in the sum of the probability (\ref{prob_spa}) for the corresponding spatial postselection. This is done with probability $P_H$. After that, the detection in the point $(0,z)$ enables the sum in the detection probability density to be independent of $\varphi_j(x,z)$, as can be seen in Eq.~(\ref{density_p}). Therefore, after normalizing the count rates we obtain the relation (\ref{prob_equal}).

Although probabilistic, our method requires neither the displacement of the detection system nor compensation in the measured count rates, since the detector will always be kept fixed at the maximum of the diffraction envelope in the point $(0,z)$ \cite{Comment}. In addition, the possibility of using programmable LCDs can result in a considerable reduction in experiment time, since instead of positioning the detector for each measurement, we have only to update the predefined configuration $(\theta_\ell,\phi_\ell)$ in the LCD and record the counts for each measurement.

\subsection{Extension to high-dimensional qudits}
The extension of this method to spatial qudits is straightforward. First, assume that in all the previous equations where a sum appears, it runs from 1 to $D$. Then, in the unitary transformation (\ref{unitary}) we set the polarization rotations and phase shifters behind each slit such that
\begin{equation}
e^{i\phi_{j\ell}}\cos(2\theta_{j\ell})=\varphi_{j\ell}^*,   \;\;\; \forall\; j,\ell=1,\ldots, D
\end{equation}
where $j$ labels the slit and $\ell$ the measurement. After measuring, in the same way as before, $D$ sets $(\theta_{j\ell},\phi_{j\ell})$ corresponding to $D$ orthogonal spatial projectors and normalizing the data we achieve the relation (\ref{prob_equal}).

\section{Conclusion}  \label{sec_conc}
In summary, we have presented an alternative measurement strategy for spatial qubits which can, probabilistically, reproduce the statistics of any observable for this system. Unlike spatial postselection \cite{Taguchi08}, our method requires neither moving the detection system nor compensation in the count rates. In addition, its extension for measuring high-dimensional qudits is straightforward. The more challenging step of the strategy would comprise the realization of the POVM, but, as suggested here, this could be achieved with the help of programmable multipixel liquid-crystal displays. We anticipate that this strategy can be a valuable tool in a variety of optical implementations of quantum information protocols with spatial qubits and qudits such as, for instance, remote state preparation, quantum state discrimination, and entanglement concentration. This will be investigated in future works.

\section*{ACKNOWLEDGMENTS}
We thank G. Lima and W. A. T. Nogueira for useful discussions. 
This work was supported by Grant Nos. CONICYT PFB08-024, PBCT Red21 and FONDECYT 11085057.


\begin{thebibliography}{99}
\bibitem{Comment2} One could also use a polarizing beam splitter rather than a polarizer and measure at the two outputs $H$ and $V$. For qubits, the statistics obtained at both outputs would reproduce that of an observable. For qudits, however, this would not be the case since there are just two outputs and $D(>2)$ outcomes. Also, it is easy to find examples where the count rates in the two outputs do not correspond to orthogonal projectors.
\bibitem{Comment} The choice of $z$ is free, though it should be far from the image plane, since around there the light attenuation is very strong at $x=0$.
\end{thebibliography}
\end{document}